**Forecasting financial distress in dynamic environments: AI adoption signals and temporally pruned training windows**

**Frederik Rech**
School of Economics, Beijing Institute of Technology,
Beijing, China
Faculty of Economics, Shenzhen MSU-BIT University,
Shenzhen, China
frederikrech@bit.edu.cn

**Hussam Musa**
Faculty of Economics, Matej Bel University,
Banská Bystrica, Slovakia
hussam.musa@umb.sk

**Martin Šebeňa**
Faculty of Arts and Social Sciences, Hong Kong Baptist University,
Hong Kong SAR, China
sebena@hkbu.edu.hk

**Siele Jean Tuo***
Business School, Liaoning University,
Shenyang, China
tuosiele88@gmail.com

## ABSTRACT

Forecasting corporate financial distress increasingly requires capturing firms' adoption of transformative technologies such as artificial intelligence, yet model performance remains vulnerable to temporal distribution shifts as these technologies diffuse. This study investigates whether firm-level artificial intelligence (AI) adoption proxies improve forecasting performance beyond standard accounting fundamentals. Using a panel of Chinese A-share non-financial firms from 2007 to 2023, we construct AI indicators from textual disclosures and patent data. We benchmark six machine learning classifiers under a strictly chronological design that fixes the final test year and progressively prunes the training history to capture temporal change. Results indicate that AI proxies consistently improve out-of-sample discrimination and reduce Type II errors, with the strongest gains in tree-based ensembles. Predictive performance is non-monotonic in training window length; models trained on recent data outperform those using full history, while single-year training proves unreliable. Explainability analyses reveal financial ratios as primary drivers, with AI adoption signals adding incremental forecasting content whose interpretation as a risk factor varies across training regimes. Our findings establish AI proxies as valuable predictors for distress screening and demonstrate that adaptive, temporally-pruned forecasting windows are essential for robust early-warning models in rapidly evolving technological and economic environments.

**Keywords:** Financial distress forecasting, Prediction models, Artificial intelligence, Machine Learning

## 1. Introduction

The accurate forecasting of corporate financial distress, particularly within rapidly evolving technological and economic environments, is a critical challenge for stakeholders seeking early warnings of firm instability. Companies facing financial distress may default on credit obligations, enter debt restructuring processes, or even declare bankruptcy due to adverse financial conditions (Song et al., 2024). While the classical literature relies on accounting ratios and market-based indicators (Altman, 1968; Beaver, 1966; Ohlson, 1980), a growing body of work shows that modern machine-learning (ML) approaches extracting more signal from complex, high-dimensional data, improving early-warning models out-of-sample predictions (Jabeur et al., 2021; Nguyen et al., 2023, Papík and Papíková, 2025). Building on these advances, recent work demonstrates that non-financial variables carry significant predictive weight alongside traditional financial metrics. Researchers incorporate macroeconomic conditions (Acosta-González et al., 2019; Sousa et al., 2022), industry structure (Sigrist and Leuenberger, 2023), regulatory environments (Fernández-Gámez et al., 2020), intellectual capital (Papíková and Papík, 2023), corporate governance (Meng et al., 2024), semantic feature in patent text (Jiang, et al., 2023), and characteristics of firms' financial information environment and information producers (Bro de Comères, 2025). These variables carry additional information about companies that is not captured in traditional financial data, and in many cases this information is inherently forward-looking.

In today's fast-paced markets, firms must innovate continuously or risk rapid erosion of their competitiveness. Yet standard accounting and market metrics understate innovation because internally generated intellectual property is weakly recognized, and investors often underreact to R&D and patent disclosures (Bai and Tian, 2020). Artificial intelligence (AI) serves as a salient case, as a general-purpose technology whose payoffs follow substantial complementary and mostly intangible investments in data, talent, processes, and organizational change, yielding a productivity J-curve in which costs and capital outlays rise before measured productivity improves (Brynjolfsson, Rock, and Syverson, 2021). In this study, firm-level AI adoption is not observed directly and is therefore measured using proxy indicators constructed from corporate disclosures and AI-related patents. China is an apt setting, as the 2017 New Generation AI Development Plan spurred rapid, policy-driven corporate adoption (Roberts et al., 2021), enabling analysis of early effects on financial health. Adoption has been shaped as much by policy incentives as by return on investment, with pronounced provincial heterogeneity and pilot zones in advanced regions like Guangdong and Zhejiang (Khanal et al., 2024); this top-down push can induce symbolic or compliance-driven uptake, including re-branding behavior, which may decouple observed AI activity from fundamentals even as authorities promote AI for financial risk monitoring and may channel support to adopters.

Conceptually, AI adoption can strengthen financial stability through three channels: (i) automation and predictive analytics that sharpen logistics, inventory, and demand forecasting, lowering cost volatility and improving working capital (Goldfarb et al., 2023); (ii) anomaly detection and AI-enabled controls that enhance reporting quality and curb governance-related distress (Chen et al., 2024; Habbal et al., 2024); and (iii) advanced analytics that raise decision quality and enable timely restructuring (Schrage et al., 2023). On the downside, the compute- and data-intensive nature of modern AI can raise a firm's carbon and resource footprint; opaque models and skewed data can entrench bias and harm; and surveillance-heavy applications may erode privacy and legitimacy, creating novel compliance and reputational exposures (Dwivedi et al., 2021). These risks are greatest when AI is treated as a plug-and-play substitute for labor; responsible adoption should emphasize human augmentation and robust governance. At the same time, AI proxies can reflect short-run transition costs and organizational disruption, and they can also reflect disclosure incentives rather than capability.

The aim of this study is to investigate whether firm-level AI adoption improves the predictive performance of corporate financial distress early-warning models beyond traditional accounting measures. We evaluate this question across multiple model families, benchmarking six widely used classifiers under a common design. Our contribution is threefold. First, we add to the growing use of non-financial features by introducing AI adoption measures into early-warning models (Papíková and Papík, 2023; Meng et al., 2024; Song et al., 2024). Second, we contribute to the literature on AI's economic impacts by extending it to the domain of corporate financial distress early-warning (Brynjolfsson et al., 2021; Goldfarb et al., 2023). Third, we use a temporally pruned training window with a fixed test year that progressively drops the oldest training year. This design is motivated by the sparse prevalence of AI proxies in early years and the rapid diffusion of AI reporting later. It allows us to study how the choice between long and short training histories affects out-of-sample early-warning performance under temporal change while preserving chronological validity. The remainder of the study is organized as follows: Section 2 details the data and methodology, Section 3 reports the empirical results, and Section 4 concludes.

## 2. Literature Review

### 2.1 Early-Warning of Corporate Financial Distress

Predicting corporate financial distress has moved from simple ratio-based discriminant models to dynamic market informed and machine learning systems. Early work showed that a few accounting ratios carry strong signals. Altman combined profitability, leverage, liquidity and activity ratios into the Z score to identify bankrupt firms (Altman, 1968) and related ratio models followed (Altman, 1983; Altman et al., 2017). Ohlson used logistic regression on ratios and firm size to relax discriminant analysis assumptions, but these static one period models often miss market signals and macro conditions (Ohlson, 1980).

Later research incorporated time and market information. Shumway (2001) pioneered a hazard model framework that treats bankruptcy as a time-varying probability, using covariates like stock returns and volatility to capture early distress signals. Campbell et al. (2008) advanced this with a dynamic logit model combining accounting and market-based variables, which significantly improved predictive accuracy and provided a useful measure for asset pricing. Contemporary refinements include incorporating macroeconomic variables (Acosta-González et al., 2019; Sousa et al., 2022), using penalized variable-selection methods in distress prediction (Li et al., 2021), and using Bayesian methods to update models with new information (Traczynski, 2017).

More recently, machine learning and AI methods ingest high dimensional financial and nonfinancial features and typically outperform traditional models out of sample (Zhong and Wang, 2022). A key challenge with black-box machine learning models in finance is their lack of interpretability, which hinders trust and regulatory adoption. To solve this, recent research emphasizes explainable AI (XAI) such as SHAP analysis to identify specific distress drivers (Deng et al., 2025; Ma et al., 2023).

At the same time, traditional static distress models tend to degrade as economic conditions evolve, since a model trained on a fixed historical sample becomes less suitable in a changing environment (Sun and Li, 2011). Major regime shifts in the economy can alter default patterns, for example, the COVID-19 shock fundamentally changed credit risk dynamics, making past relationships unreliable without proper temporal validation (Breeden, 2025). Empirical studies confirm that concept drift in distress data is real and that using rolling training windows can significantly improve predictive performance compared to static estimation (Sun and Li, 2011). Therefore, careful choice of the training window and rigorous out-of-time validation are crucial to maintain robust accuracy under time variation and distribution shifts. These temporal dynamics are particularly relevant when evaluating emerging, unevenly distributed predictors like firm-level AI adoption, which we explore next.

### 2.2 Artificial Intelligence and financial distress

Firms adopt AI to enhance performance by automating processes, enabling data-driven decisions, and generating predictive insights. Researchers have employed diverse methodologies to capture firm-level AI adoption, each with distinct advantages and limitations. Common approaches include textual analysis of corporate disclosures (Li et al., 2025), employee skill indices derived from résumés (Babina et al., 2024), AI-related job postings (Han et al., 2025), and industry-level robot penetration (Liu et al., 2025). This section analyzes how these AI capabilities influence financial distress by enhancing operational efficiency, fostering innovation, and strengthening risk management. These functions represent key mechanisms through which AI may help firms anticipate or avoid financial distress.

AI adoption significantly influences firm performance and risk through several channels. AI drives product innovation and growth, with firms investing in AI experiencing higher sales, employment, and market value growth, primarily through new product development (Babina et al., 2024). Recent evidence from China shows that firm-level AI engagement is associated with firms' ability to balance short-term returns and long-term growth, with effects varying across firm characteristics and development stages (Lin and Sun, 2026), reinforcing the view that observed AI signals reflect heterogeneous strategic positioning rather than uniform outcomes. Consistent with this interpretation, AI adoption is associated with differences in how firms perceive and manage financial and operational risks, suggesting that observed AI signals may capture broader risk-management orientation rather than immediate performance effects (Dvorsky, 2025). However, evidence on operational efficiency is mixed. While AI is theorized to enhance operational efficiency through automation and optimized decision-making (Dwivedi et al., 2021), empirical evidence remains mixed. Some studies demonstrate significant productivity gains, such as a 14% increase in output from generative AI assistance (Brynjolfsson et al., 2023). However, these gains are not universal, as other research finds that efficiency improvements are heavily contingent on complementary factors like data infrastructure (Bessen et al., 2022). Furthermore, AI's impact manifests through complex changes in labor composition rather than simple automation (Acemoglu et al., 2022), suggesting that net efficiency gains may be muted in the short run as firms undergo this transition. This aligns with the 'productivity J-curve' (Brynjolfsson et al., 2019), where initial investment costs outweigh benefits before eventual gains are realized. On the other hand, Li et al. (2025) find that AI shows no association with improvements in gross operational efficiency and is actually negatively related to net operational efficiency in Chinese-listed firms.

Methodologically, researchers have employed various identification strategies to establish causality. Some studies exploit policy shocks or industry trends as instruments for AI investment, such as regional AI initiative rollouts, to isolate causal impacts on performance. A consistent finding across these studies is that firm heterogeneity mediates AI's benefits. Larger firms and those with complementary intangibles tend to reap greater performance gains (Babina et al., 2024; Song et al., 2022), while smaller or resource-constrained firms often adopt AI superficially, yielding limited improvements. Many firms appear stuck in a "pilot trap" characterized by experimenting with AI in isolated processes but not achieving scale, which limits measurable performance impact. This underscores the need for complementary investments in training and reorganization to unlock AI's potential, consistent with the "productivity J-curve" theory (Brynjolfsson et al., 2021).

A growing body of literature also examines AI's role in enhancing corporate risk management and resilience. Firms with AI-skilled workers recovered nearly all market value after natural disasters, attributed to superior operational adaptability (Han et al., 2025). Liu et al. (2025) demonstrate that AI reduces financial risks by enhancing talent recruitment, alleviating financing constraints through a resource effect, and improving total factor productivity. Paradoxically, AI adoption may also encourage greater strategic risk-taking. Chen et al. (2024) document that firms deeply integrating AI show approximately 6.5% greater risk propensity in investments and financing decisions, possibly because AI's predictive capabilities enable firms to pursue opportunities more aggressively.

## 3. Research Methodology

### 3.1. Data sources and samples

Our research constructs a 2007–2023 firm-year panel for Chinese A-share companies using annual reports from Sina Finance, AI-related patent data from Chinese Research Data Services (CNRDS) Platform, and financial-statement variables from CSMAR. We remove financial-sector observations and firms in information transmission/software/ IT services and scientific research/technical services, where cloud, big data, and AI are already standard practice and heavily disclosed, preventing clear identification of AI adoption on its own. The merged sample comprises 33,097 firm-years, including 1,041 (3.05%) distressed observations. To address missing data, we applied multiple imputation by chained equations (mice) with five imputations, using CART-based conditional models.

Distress is identified via the China Securities Regulatory Commission's (CSRC) Special Treatment (ST) or *Special Treatment (*ST) designation (Huang et al., 2024; Meng et al., 2024). Most ST/*ST cases arise from two consecutive years of losses (about 80%); other cited triggers include operating losses, investment risk, bankruptcy, shrinking assets or equity contraction, and temporary pauses in operations (Meng et al., 2024). Following common practice, we predict distress at year t using each firm's ST/*ST status measured at t– 2 (Song et al., 2024). We define healthy firms as those not flagged ST or *ST at any time in our sample. For distressed outcomes, we drop cases where the base-year observation (t−2) is already ST/*ST, so the model does not simply project an existing distressed status two years forward.

### 3.2 Variable Selection

Table 1 summarizes the variables used in our predictive models. The feature set comprises the five Altman Z-score components (Altman, 1968) with a suite of proxies designed to capture a firm's engagement with AI technologies. All AI adoption variables are constructed as indirect proxies, as true AI integration is not directly observable in corporate disclosures.

Measuring firm-level AI adoption is challenging due to its diverse applications, rapid evolution, and integration into broader digital processes. While survey-based metrics are limited in scale, text-based indicators from corporate disclosures offer a scalable and a popular alternative (Han and Meng, 2025; Chen et al., 2025; Yang and Yang (2025). Following this approach, we construct a composite measure based on a unified 72-term AI lexicon compiled from prior academic lists, national research reports, and the WIPO AI glossary, harmonize Chinese and English variants, and collapse synonym. Using natural language processing techniques, we identify and aggregate occurrences of AI-related keywords across both documents, including core terms such as "artificial intelligence", "machine learning", "natural language processing", "Internet of Things", "distributed computing", and "big data analytics". Our primary firm-year AI adoption metric is the natural logarithm of one plus the total count of these terms across all disclosures. Similarly, we identify AI patents by applying the same lexicon to patent titles and abstracts, creating analogous measures for AI invention, utility model, and design patents (Chen and Zhang, 2025).

**Table 1**
Variables Definitions.

| Variable | Calculation | Definition |
|---|---|---|
| Financial variables | | |
| X01 | WorkingCapital / TotalAssets | Short- term liquidity; higher values indicate stronger ability to meet current obligations |
| X02 | RetainedEarnings / TotalAssets | Accumulated profitability; reflects internal financing capacity and loss- absorption cushion |
| X03 | EBIT / TotalAssets | Operating performance efficiency; gauges how effectively assets generate operating profit |
| X04 | MarketValueOfEquity / TotalLiabilities | Market- based leverage; higher values imply a larger equity buffer relative to debt |
| X05 | OperatingRevenue / TotalAssets | Asset productivity; measures sales generated per unit of assets |
| AI variables | | |
| AI level | Ln (1+AI frequency in AR) | Captures the frequency of AI-related terms in the entire annual report |
| AI level MD&A | Ln (1+AI frequency in MD&A) | Captures the frequency of AI-related terms in the Management Discussion & Analysis (MD&A) section |
| AI patents | Ln (1 + [AI invention + AI utility + AI design]) | Total AI patent applications (invention + utility model + design; independent + joint). |
| AI invention | Ln (1 + AI invention) | AI invention patent applications (independent + joint; novel technical solutions). |
| AI utility | Ln (1 + AI utility) | AI utility model patent applications (independent + joint; functional improvements). |
| AI design | Ln (1 +AI design) | AI design patent applications (independent + joint; aesthetic/ornamental design). |

Figure 1 tracks how often each variable is nonzero over time for both healthy and distressed firms. In both panels the early years 2009–2014 are dominated by zeros. Almost no firms register activity on the AI measures or the innovation proxies, so usable observations are scarce. Coverage climbs from 2015, especially broad AI adoption, and by 2020–2021 only about half of companies show any AI activity.

Healthy firms generally show higher and earlier uptake. Their AI adoption rises smoothly year after year and ends the period a bit above the distressed sample. Distressed firms also catch up late in the horizon, but their levels are slightly lower, and the paths look bumpier (also see Fig. 2). Distressed firms catch up late but remain lower and bumpier (also see Fig. 2).

Auxiliary innovation signals such as patents, invention, utility, or design filings stay sparse (single digits to low teens), with brief late 2010s spikes, especially among healthy firms.

China's State Council's 2017 New Generation AI Development Plan and provincial subsidies and targets created strong incentives for AI-related patenting, helping set up a visible rise in 2019–2020 (Roberts et al., 2021). The spike can be explained by the 18-month publication lag because a 2018 filing wave would appear in 2019–2020 and mechanically lift counts (Okada and Nagaoka, 2020). Moreover, the dominance of invention patents is to be expected because subsidy amounts for invention patents are significantly higher than for utility models or design patents, further reinforcing the observed pattern (Dang and Motohashi, 2015).

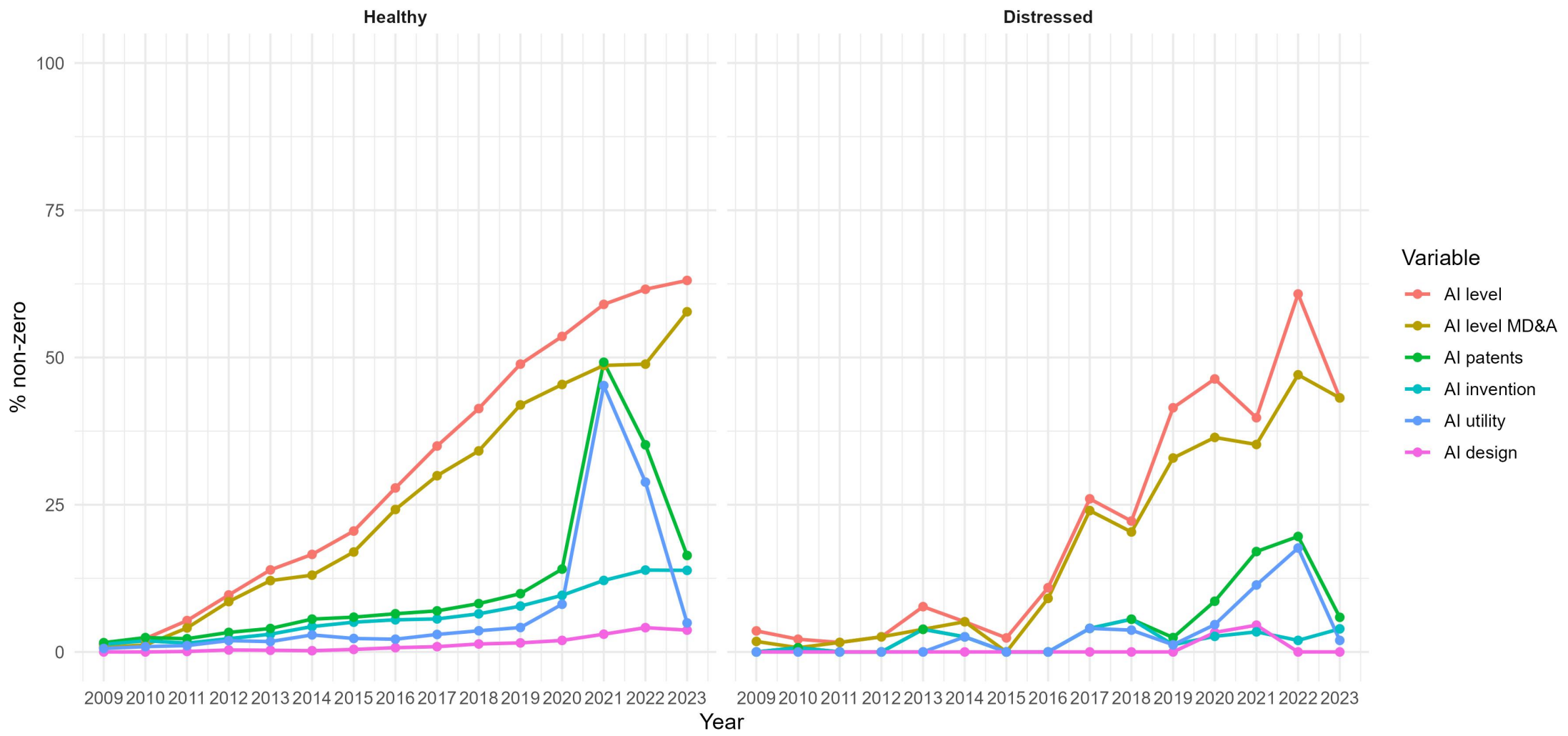

**Fig. 1.** Prevalence of AI measures over time (% of firms).

In line with Figure 1, the cross-sectional means show healthy firms scoring higher on every measure, including AI adoption levels, AI term densities, and patent activity across invention, utility, and design classes. The gap suggests healthier firms are more active in AI adoption on average.

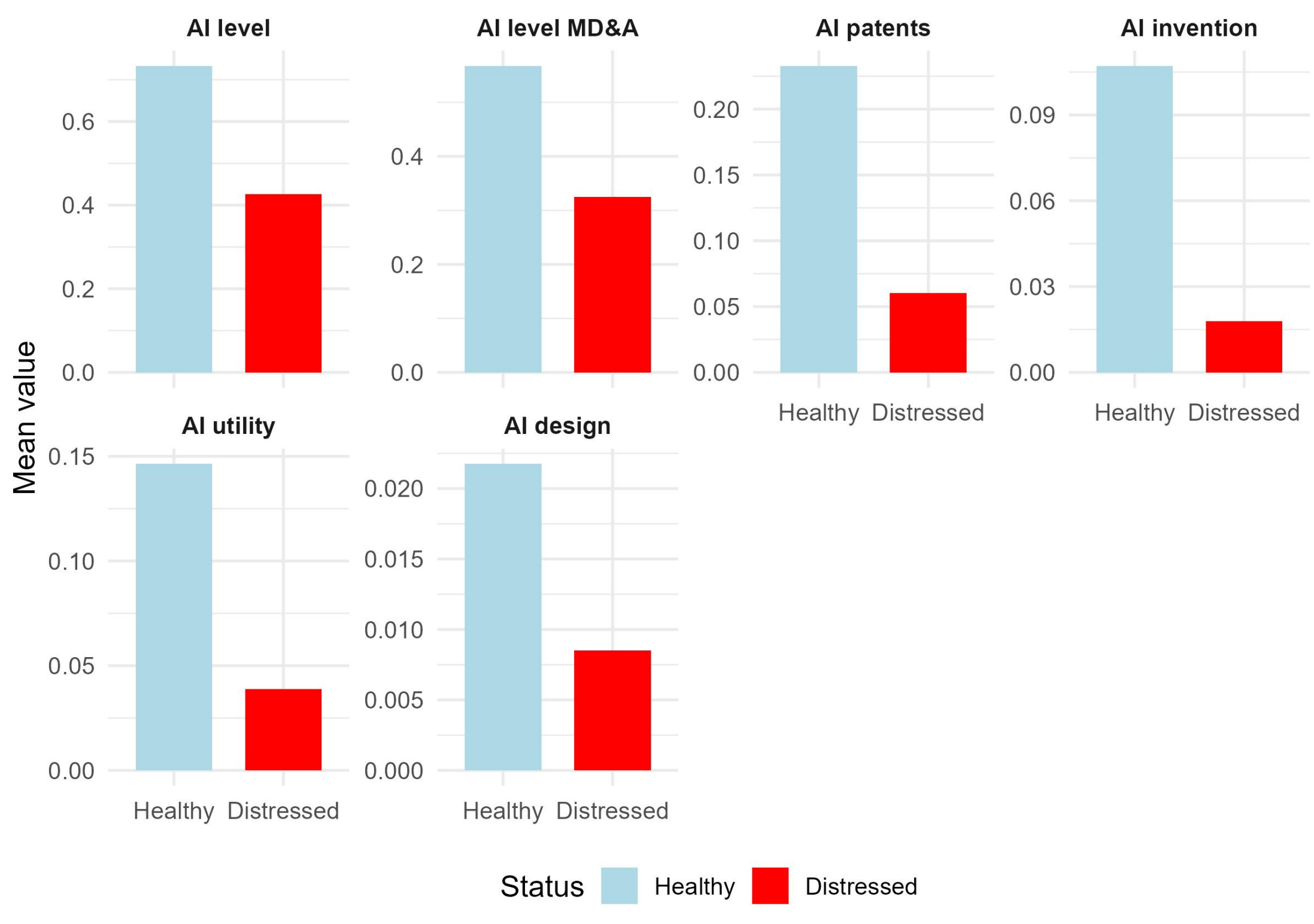

**Fig. 2**. Mean AI adoption and innovation intensity by financial health.

### 3.3. Model construction and evaluation

Prior studies often evaluate models in ways that can overstate out-of-sample performance. Several rely mainly on in-sample assessments that do not enforce chronological separation, yielding optimistic fit statistics with little ex-ante content (Geng et al., 2015; Song et al., 2024). Others do perform ex-ante model validation, but starve the models with low count of data, either because of the data sample or under-sampling strategies (Geng et al., 2015; Ma et al., 2023; Meng et al., 2024). A smaller subset of studies addresses these challenges by using instance selection or dynamic pruning to mitigate concept drift and better align training data with future test periods (Sun and Li, 2011), consistent with evidence that predictor relevance shifts under external shocks and late-stage distress dynamics (Rech et al., 2025). These issues are amplified in our setting, as AI adoption is scarce in the early sample (Fig. 1), reflecting an emerging technology; in subsequent years, adoption becomes more common, reducing the share of zeros and altering the underlying distribution.

We therefore implement a pruned training window with a fixed test year (Fig. 3). The test set is 2023 and remains untouched during training and tuning. For split $k = 1, \ldots, 14$, we train on $[s_k, 2022]$ and evaluate once on 2023, where $s_k \in \{2009, \ldots, 2021\}$. Thus, the shortest window uses one year (2022); the longest uses the whole sample (2009–2022). This produces 14 chronologically valid training samples that shrink in length:

$$T_k = [s_k, 2022], \qquad S = \{2023\},$$

The design mimics real-time forecasting avoids look-ahead bias and lets us study how performance evolves as more data is available and enters the sample.

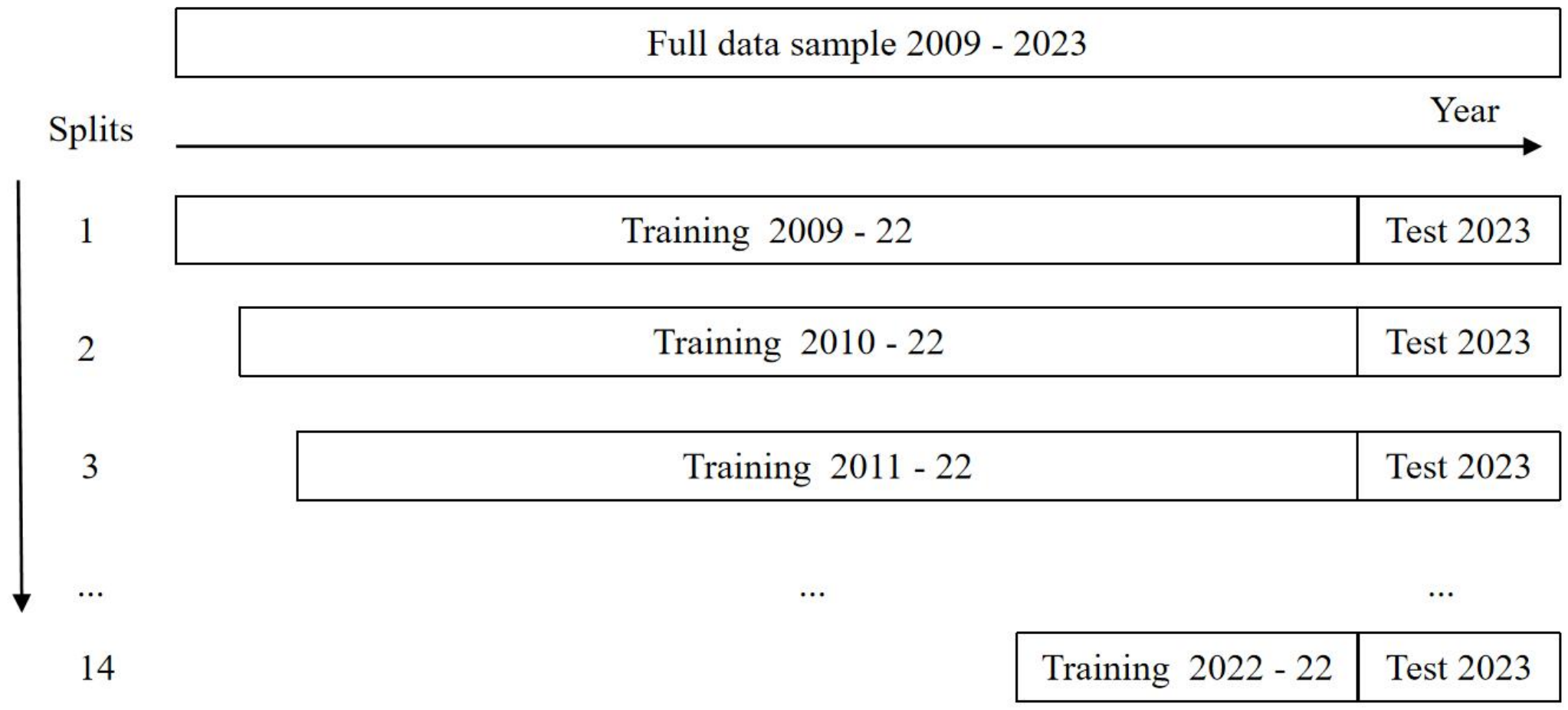

**Fig. 3** Pruned training window with fixed test year.

Within each training window we winsorize each predictor at the 1st/99th percentiles computed from the training slice only, then apply those limits to the test set. With the same procedure we also apply Z-score standardization. Within split $s$, we compute:

$$x^s_{z-score} = \frac{x - \mu^s_{train}}{\sigma^s_{train}},$$

where $\mu^s_{train}$ and $\sigma^s_{train}$ are the mean and standard deviations of the training samples.

We model two scenarios, one that augments financial variables with AI features and another that relies on financial variables alone. We estimate six classifiers, namely XGBoost, LightGBM, Random Forest (RF), Logistic Regression, Neural Networks (NN), and radial basis function support vector machine (RBF-SVM). For each $T_k$, hyperparameters are selected by 10-fold stratified cross-validation inside the training window only; the 2023 test data are never used for preprocessing, fitting, tuning, or threshold selection.

Class imbalance is handled via inverse-frequency class weights. Let $n_1$ and $n_0$ be the counts of distressed and healthy firm-years in $T_k$. We preserve the event rate across folds (class-stratified) and optimize a class-weighted loss with inverse-frequency weights:

$$w_1 = \frac{n}{2n_1}, w_0 = \frac{n}{2n_0}, \qquad L = \sum_i w_{y_i} \ell(y_i, \quad \hat{p}_i),$$

so the minority (distress) cases receive increased loss weight without synthetic rebalancing. This keeps the AI = 0 distribution intact and avoids starving the model of majority-class information that is essential for calibration.

We evaluate on a held-out 2023 test set, reporting AUC, Accuracy, Recall, Specificity, Precision, F1, G-Mean, and confusion-matrix counts as both split averages and split-specific results. For interpretation, we use SHAP explanations alongside model-specific feature importance.

## 4. Empirical results

### 4.1. Main results

Table 2 presents the out-of-sample performance of our six machine learning models, evaluated with and without the augmentation of AI-derived features, while Table 3 provides a statistical analysis of the effect of adding these AI variables across multiple data splits. The numbers in bold in Table 2 indicate the superior model within each algorithm pair (with vs. without AI).

**Table 2**
Model Performance Comparison (With vs. Without AI)

| | AUC | Accuracy | Recall | Specificity | F1-Score | G-Mean | TP | TN | FP | FN |
|---|---|---|---|---|---|---|---|---|---|---|
| XGB — with AI | **0.867** | **0.854** | **0.714** | **0.856** | **0.115** | **0.781** | **36.43** | **3279.71** | **553.29** | **14.57** |
| XGB — without AI | 0.860 | 0.838 | 0.686 | 0.840 | 0.103 | 0.757 | 35.00 | 3218.07 | 614.93 | 16.00 |
| LightGBM— with AI | **0.876** | 0.857 | 0.702 | **0.859** | **0.121** | 0.769 | 35.79 | **3293.21** | **539.79** | 15.21 |
| LightGBM— without AI | 0.870 | 0.851 | 0.711 | 0.853 | 0.114 | 0.776 | 36.29 | 3270.14 | 562.86 | 14.71 |
| RF— with AI | **0.881** | **0.854** | **0.744** | **0.856** | **0.119** | **0.797** | **37.93** | **3279.86** | **553.14** | **13.07** |
| RF-— without AI | 0.864 | 0.852 | 0.697 | 0.854 | 0.111 | 0.771 | 35.57 | 3274.07 | 558.93 | 15.43 |
| LR— with AI | **0.880** | **0.869** | **0.706** | **0.871** | **0.126** | **0.783** | **36.00** | **3340.21** | **492.79** | **15.00** |
| LR— without AI | 0.875 | 0.865 | 0.703 | 0.867 | 0.122 | 0.780 | 35.86 | 3323.36 | 509.64 | 15.14 |
| NN— with AI | **0.879** | 0.835 | **0.772** | 0.836 | 0.115 | **0.802** | **39.36** | 3204.86 | 628.14 | **11.64** |
| NN— without AI | 0.870 | 0.857 | 0.732 | 0.859 | 0.119 | 0.793 | 37.36 | 3291.79 | 541.21 | 13.64 |
| SVM— with AI | **0.886** | **0.850** | **0.758** | **0.851** | **0.120** | **0.802** | **38.64** | **3263.21** | **569.79** | **12.36** |
| SVM— without AI | 0.867 | 0.847 | 0.734 | 0.849 | 0.115 | 0.789 | 37.43 | 3254.07 | 578.93 | 13.57 |

Notes: Metrics are averaged over 14 splits under a pruned training window. Split 1 trains on 2009–2022; in each subsequent split we drop the oldest training year (2009–2022 ··· 2022–2022) while the test set remains 2023, preserving its class imbalance and isolating how performance responds to the evolving training history.

The central finding is that AI variables improve discrimination across all learners. AUC rises in every model, from +0.005 in logistic regression to +0.019 in SVM. Tree-based ensembles also benefit materially, Random Forest +0.017, XGBoost +0.007, and LightGBM +0.006; the neural network improves by +0.009. Although these AUC gains are numerically small, they are economically meaningful because baseline AUC levels are already close to 0.87, where further improvements are typically difficult to obtain and can still change real classification outcomes under imbalance.

The improvements are most pronounced in detecting distress under class imbalance. Recall increases for five models, with gains of 0.028 for XGBoost, 0.047 for Random Forest, 0.003 for logistic regression, 0.040 for the neural network, and 0.024 for SVM, while LightGBM declines slightly by 0.009. G-Mean shows a consistent pattern, improving by 0.024 for XGBoost, 0.026 for Random Forest, 0.003 for logistic regression, 0.009 for the neural network, and 0.013 for SVM, and falling by 0.007 for LightGBM. Confusion-matrix components indicate that gains are primarily driven by fewer missed failures. Random Forest increases true positives by 2.36 and reduces false negatives by 2.36. XGBoost increases true positives by 1.43 and reduces false negatives by 1.43. SVM increases true positives by 1.21 and reduces false negatives by 1.21. Put differently, the AI variables help capture roughly 1 to 3 additional distressed firms on average, depending on the learner. In decision terms, this matters because each additional correctly flagged distressed firm can avoid high-cost errors linked to missed failures, such as avoidable credit losses, delayed intervention, or underestimation of portfolio risk, whereas the cost of additional screening from false positives is often comparatively lower. The neural network is the main exception in terms of trade-offs, with higher recall but substantially more false positives and lower accuracy.

**Table 3**
Effect of AI Adoption Features Across Splits.

| | Effect of adding AI variables | | Inference | | | Direction |
|---|---|---|---|---|---|---|
| | Δ (With − Without) | 95% CI | t statistic | p-value (t) | p-value (boot) | |
| AUC | | | | | | |
| XGBoost | 0.0068 | [ 0.0040, 0.0094] | 4.8757 | 0.0000*** | 0.0000*** | ↑ AI better |
| LightGBM | 0.0045 | [ 0.0009, 0.0080] | 2.3768 | 0.0226** | 0.0172** | ↑ AI better |
| RF | 0.0072 | [ 0.0040, 0.0104] | 4.3456 | 0.0001*** | 0.0000*** | ↑ AI better |
| Logit | 0.0023 | [ 0.0000, 0.0047] | 1.8619 | 0.0704* | 0.0496** | ↑ AI better |
| NN | −0.0066 | [-0.0158, 0.0027] | −1.3785 | 0.1761 | 0.1668 | ↓ AI worse |
| SVM | 0.0047 | [-0.0018, 0.0120] | 1.3364 | 0.1894 | 0.1692 | ↑ AI better |
| F1 | | | | | | |
| XGBoost | 0.0111 | [ 0.0031, 0.0197] | 2.4875 | 0.0272** | 0.0052*** | AI better |
| LightGBM | 0.0066 | [-0.0007, 0.0169] | 1.3494 | 0.2002 | 0.0948* | AI better |
| RF | 0.0083 | [ 0.0034, 0.0129] | 3.2765 | 0.0060*** | 0.0020*** | AI better |
| Logit | 0.0044 | [ 0.0020, 0.0067] | 3.4814 | 0.0041*** | 0.0008*** | AI better |
| NN | −0.0047 | [-0.0127, 0.0032] | −1.1324 | 0.2779 | 0.2408 | AI worse |
| SVM | 0.0048 | [-0.0020, 0.0110] | 1.4145 | 0.1807 | 0.1496 | AI better |
| GMean | | | | | | |
| XGBoost | 0.0239 | [ 0.0096, 0.0382] | 3.1223 | 0.0081*** | 0.0008*** | AI better |
| LightGBM | −0.0069 | [-0.0302, 0.0081] | −0.6326 | 0.5379 | 0.6224 | AI worse |
| RF | 0.0258 | [ 0.0144, 0.0379] | 4.0748 | 0.0013*** | 0.0000*** | AI better |
| Logit | 0.0032 | [-0.0067, 0.0132] | 0.6065 | 0.5546 | 0.5472 | AI better |
| NN | 0.0094 | [-0.0058, 0.0236] | 1.2242 | 0.2426 | 0.2056 | AI better |
| SVM | 0.0136 | [ 0.0030, 0.0243] | 2.4410 | 0.0297** | 0.0092*** | AI better |
| Type I error (FPR) | | | | | | |
| XGBoost | 0.0161 | [-0.0034, 0.0420] | 1.3066 | 0.2140 | 0.1536 | AI reduces Type I |
| LightGBM | 0.0060 | [-0.0028, 0.0149] | 1.2784 | 0.2235 | 0.1876 | AI reduces Type I |
| RF | 0.0015 | [-0.0118, 0.0137] | 0.2217 | 0.8280 | 0.7976 | AI reduces Type I |
| Logit | 0.0044 | [-0.0013, 0.0094] | 1.5403 | 0.1475 | 0.1088 | AI reduces Type I |
| NN | −0.0227 | [-0.0476, -0.0033] | −1.8973 | 0.0802* | 0.0156** | AI increases Type I |
| SVM | 0.0024 | [-0.0176, 0.0209] | 0.2362 | 0.8170 | 0.7836 | AI reduces Type I |
| Type II error (FNR) | | | | | | |
| XGBoost | 0.0280 | [-0.0014, 0.0574] | 1.7785 | 0.0987* | 0.0676* | AI reduces Type I |
| LightGBM | −0.0098 | [-0.0406, 0.0140] | −0.6630 | 0.5189 | 0.5620 | AI increases Type I |
| RF | 0.0462 | [ 0.0168, 0.0784] | 2.7761 | 0.0157** | 0.0016** | AI reduces Type I |
| Logit | 0.0028 | [-0.0182, 0.0238] | 0.2494 | 0.8069 | 0.8180 | AI reduces Type I |
| NN | 0.0392 | [ 0.0042, 0.0756] | 2.0455 | 0.0616* | 0.0276** | AI reduces Type I |
| SVM | 0.0238 | [ 0.0056, 0.0406] | 2.5171 | 0.0257** | 0.0136** | AI reduces Type I |

Notes: Δ for AUC/F1/GMean = WithAI − WithoutAI (↑ means AI better). Δ for Type I/II = WithoutAI − WithAI (↑ means AI reduces error). Significant differences at the 90%, 95% and 99% levels are indicated by *, ** and ***, respectively.

Table 3 shows that the effects of the AI features are broadly consistent across the pruned-window splits and are not driven by any single training history. The most statistically significant improvements are in discrimination, with clear AUC gains for XGBoost and Random Forest and a smaller but significant gain for LightGBM; logistic regression shows weaker, borderline evidence, while SVM is positive but not statistically significant. For imbalance-sensitive metrics, F1 improves for XGBoost, Random Forest, and logistic regression, and G-Mean improves for XGBoost, Random Forest, and SVM, with some additional borderline evidence in a few models depending on the inference method. Type I error effects are generally not significant, except that the neural network shows evidence of worsening false positives. Type II error declines for Random Forest and SVM, with marginal evidence for XGBoost, implying that AI variables primarily help reduce missed distress events.

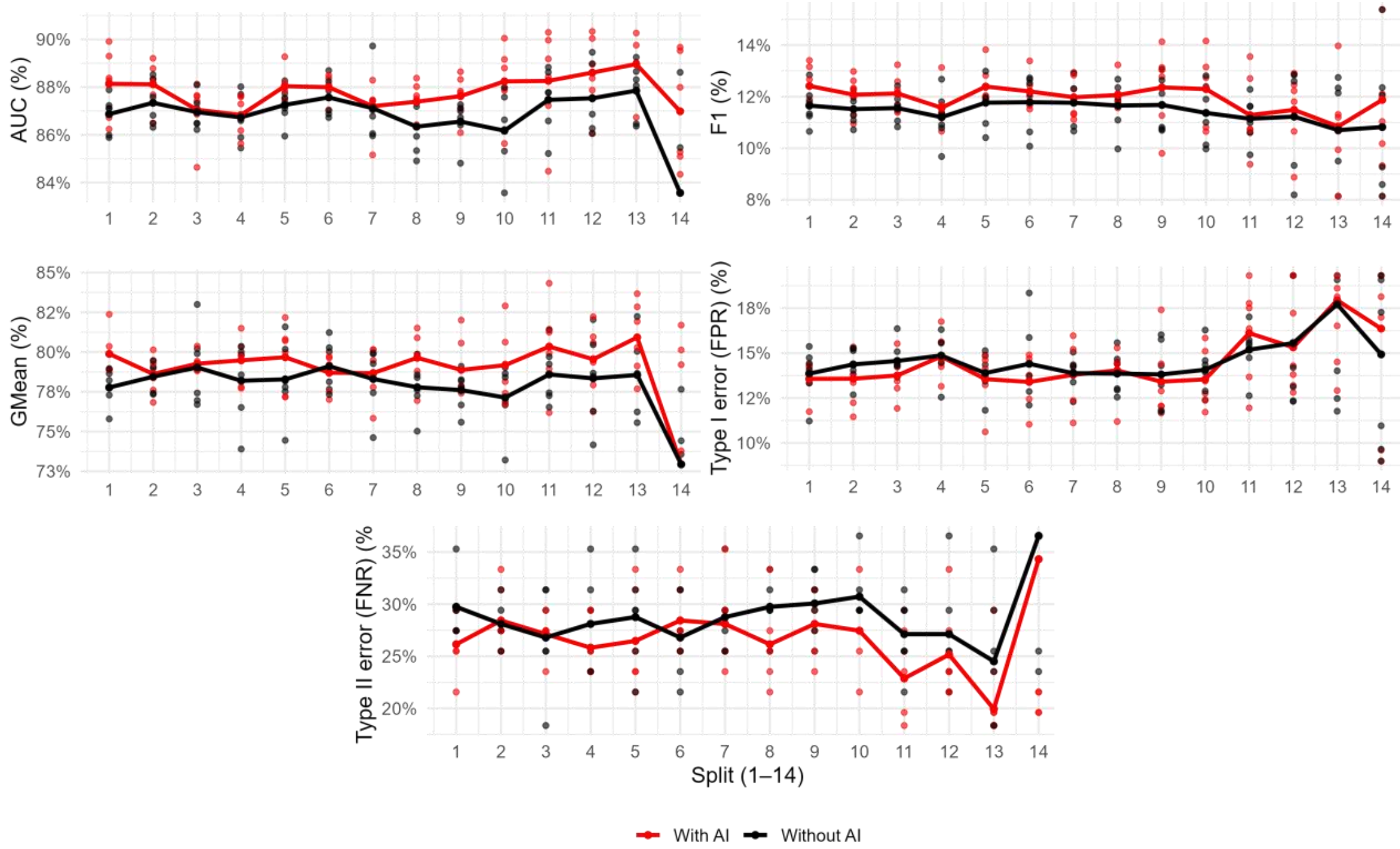

**Fig. 4.** Split wise model performance (2023 test).

Figure 4 shows split wise performance for models with and without AI features and makes clear that the optimal training window depends on the prediction objective, whether prioritizing distressed firm detection or limiting false alarms. While the longest training window is not the worst case, it is also not the best, which suggests that using all available history can dilute time varying relationships and that temporal dynamics matter even in the cohort without AI.

Surprisingly, the best performance was achieved in split 13, which trains only on two most recent years (2021 to 2022). In this split, identification of distressed firms is highest for both specifications, with the lowest Type II error and the highest AUC and GMean. This comes with a clear trade off, as Type I error rises, consistent with a more aggressive screening rule that captures more true distress cases but flags more healthy firms. This peak is not isolated. Performance starts to improve from splits 11 to 13 as the training window concentrates on recent years, and within this range the AI specification shows a more pronounced reduction in Type II error and stronger AUC and GMean than the no AI baseline, with the gap between the two series becoming more stable and then widening as the window shortens.That said, pushing the window too far is clearly detrimental. Split 14, trained only on 2022, is the worst performing and visibly unstable across metrics, indicating that a single year provides too little information for reliable learning and should not be used in practice.

**4.2 Model interpretability and economic drivers of prediction**

Figure 5 summarizes how the model attributes 2023 predictions to each predictor under alternative training histories, with feature values re-scaled using the 2009–2022 training window so that colour gradients are directly comparable across panels.

The first takeaway is that financial fundamentals remain the dominant drivers in every window and their directional patterns are largely stable. The key change is in the dispersion of their predictive contributions. As we shorten the training window, the SHAP distributions for several financial variables contract and their extreme tails thin out, most clearly for x1 and x4, which cluster closer to zero in the 2021–2022 window. From a predictive perspective, this indicates that very long training histories smooth over multiple regimes, whereas shorter windows yield more concentrated and context-relevant signal that aligns more closely with the 2023 prediction environment.

The second takeaway is that the AI measures show clearer shifts in both dispersion and sign. AI level has a negative slope in every window. High AI level observations tend to sit on the left side of the SHAP axis, so they reduce predicted distress. Low AI level observations are closer to zero or slightly positive, so the absence of AI adoption does not provide the same protection. This asymmetric pattern suggests that AI adoption acts as a downside-risk mitigator rather than a symmetric performance enhancer.

Interestingly, AI level MD&A behaves differently from the other AI measures. In the long training window, its SHAP values spread to both sides of zero, but the positive side is longer, so high MD&A AI disclosure more often coincides with higher predicted distress. In the medium window, the distribution is almost symmetric around zero, so the model does not assign a clear directional interpretation on average. In the short window, the pattern tightens and shifts to the positive side, so AI level MD&A mainly functions as a one-sided predictor of elevated distress rather than a mitigating signal.

AI patent and AI invention load negatively across windows. High patent and invention intensity sits on the negative SHAP side, corresponding to lower predicted distress, while low intensity is closer to zero. Their spreads compress sharply in the 2021 – 2022 window, indicating that this predictive signal becomes less cross-sectionally extreme once the model is trained entirely within a high AI adoption period. LnUtility remains tightly clustered near zero and mostly on the negative side, so its predictive role is weakly protective when present. Lastly, LnDesign is the one AI patent proxy that consistently shifts predictions toward higher distress across all training windows, especially in the longest and the shorter windows, suggesting that design-oriented AI activity is systematically associated with higher predicted risk in the model's ranking.

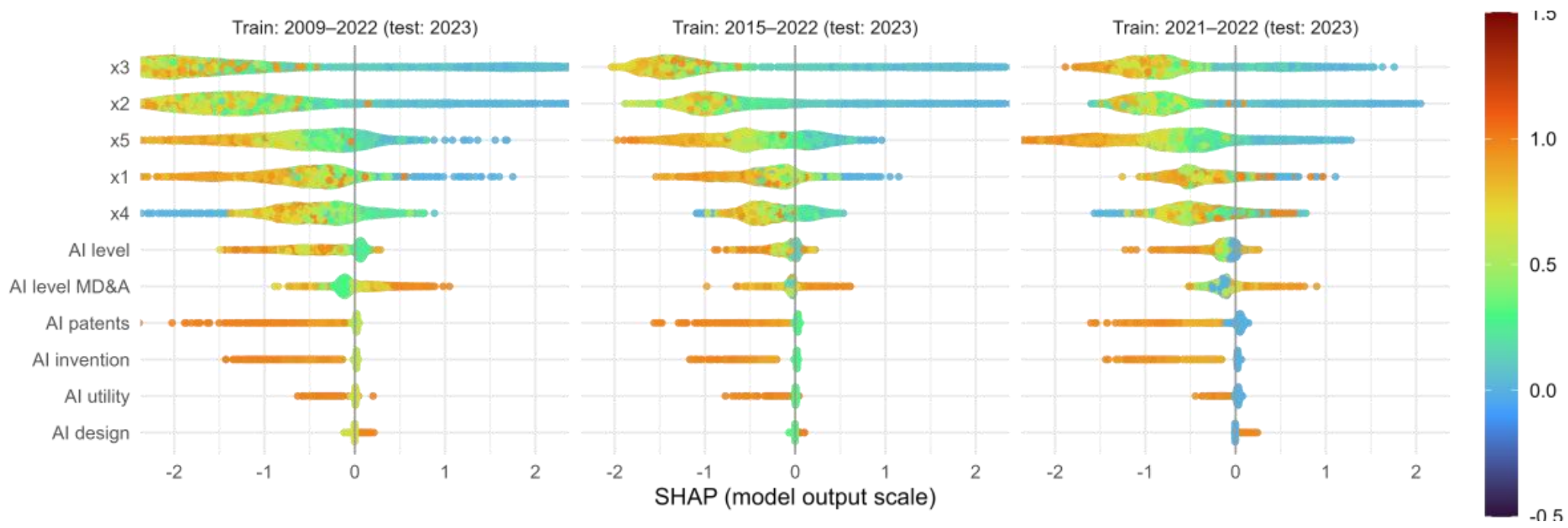

**Fig. 5** Global SHAP beeswarm across training windows

Panel a of Figure 6 tracks mean absolute SHAP contributions on the fixed 2023 test set as we move from long to short training windows. The dominant pattern is that the core financial predictors remain first order in every split, consistent with distress being primarily predicted from fundamentals. What changes with the window is how strongly the model leans on each signal. When the training window concentrates on recent years, importance becomes more time sensitive. In particular, the AI variables show a clear increase in importance in the late splits, especially around splits 11 to 13, when the training data are drawn from a period of more prevalent AI adoption. This is also where out of sample results show stronger identification of distressed firms, so the higher SHAP importance has direct practical relevance. Split 14 is different. The sharp jump in mean absolute SHAP for many variables indicates an unstable mapping from inputs to predicted risk when the model is trained on a single year, suggesting limited decision usefulness under extremely short histories.

The uncertainty view in Panel b sharpens the interpretation. Financial drivers show clear separation in levels, while the ranking changes between x2 and x3 in the late splits reflect a shift in predictive weight rather than noise. Within the AI block, AI level and AI level MD&A are both higher and more precisely estimated than AI patents and AI Invention. In predictive terms, the model relies more on contemporaneous adoption and disclosure signals than on slower moving patent stocks, which add incremental information but rarely drive rankings on their own.

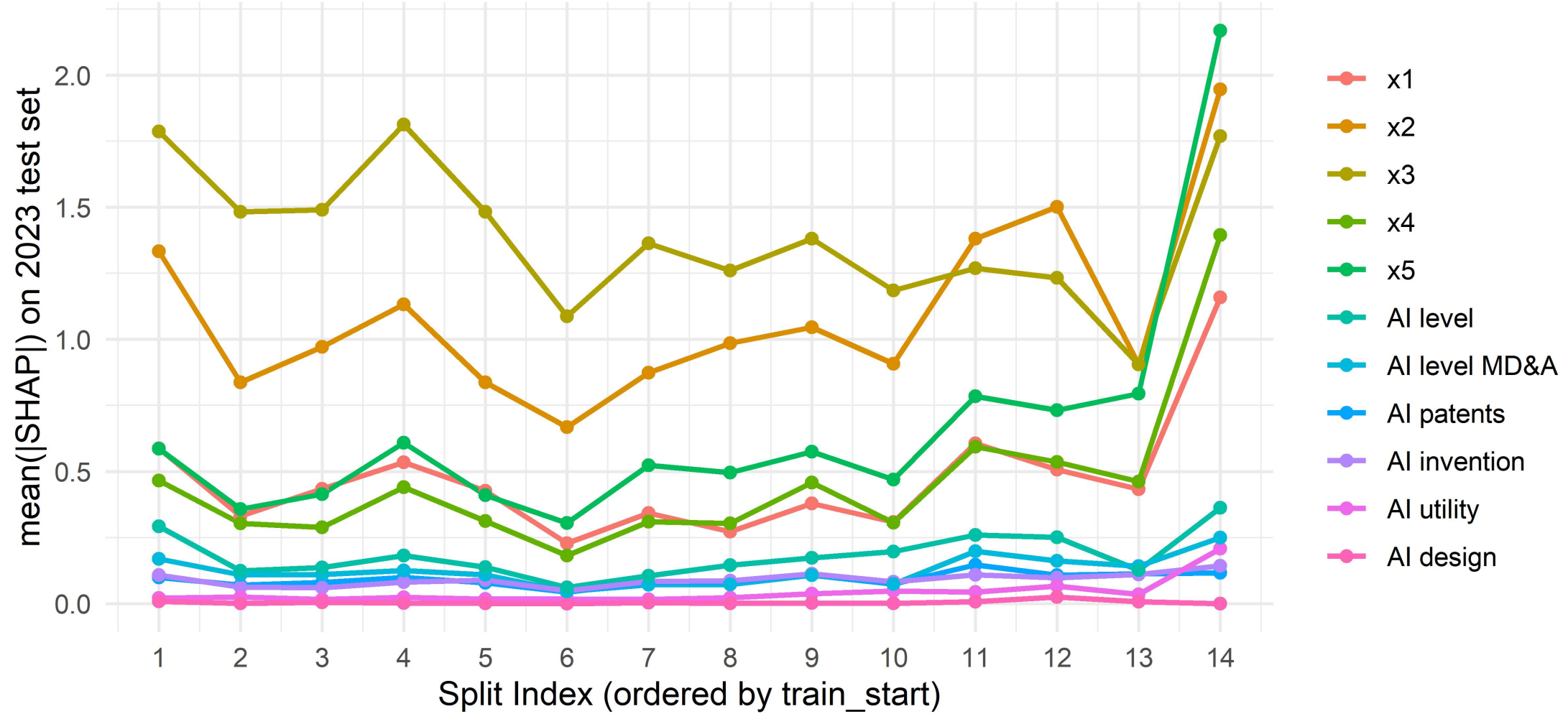

(a) Importance vs split

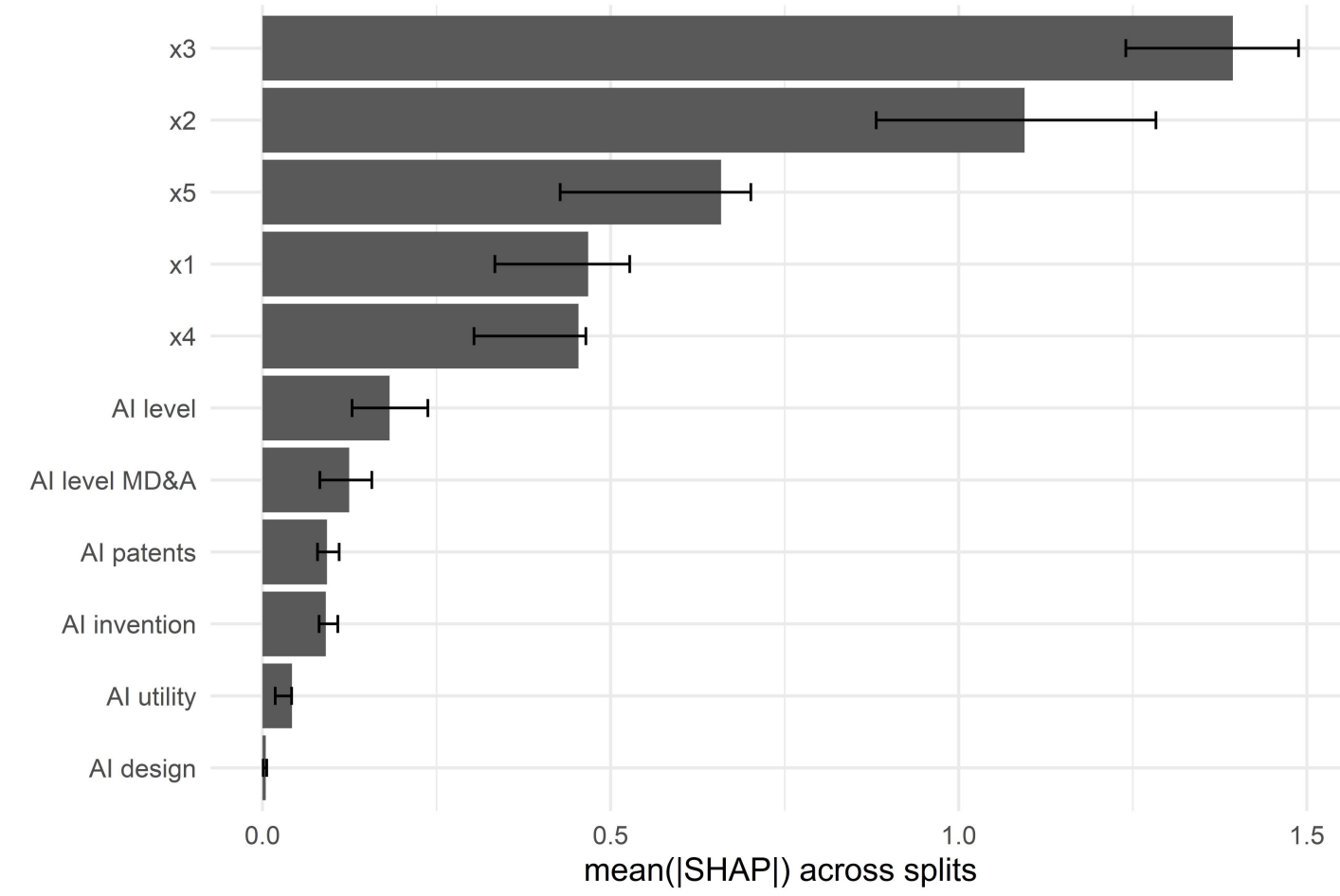

(b) Global importance with uncertainty

**Fig. 6** SHAP importance over pruned windows and uncertainty

Figure 7 reports the split wise composition of XGBoost gain, shown as one stacked bar per training window, so each bar sums to 100 percent and the colours indicate which variables account for the model's split driven improvement in fit as the window shortens from 2009–2022 to 2022 only. Unlike SHAP, which decomposes each 2023 prediction into signed feature contributions, gain is an unsigned training time importance measure that reflects how much each variable reduces the objective when used for splits and it can shift with collinearity even when predictive content is similar. The gain composition plot broadly confirms the SHAP pattern. Financial ratios dominate in every window. The AI proxies remain smaller, but they gain visible share in the late splits, especially the disclosure based measures AI level and AI level MD&A, with AI patent intensity contributing at the margin.

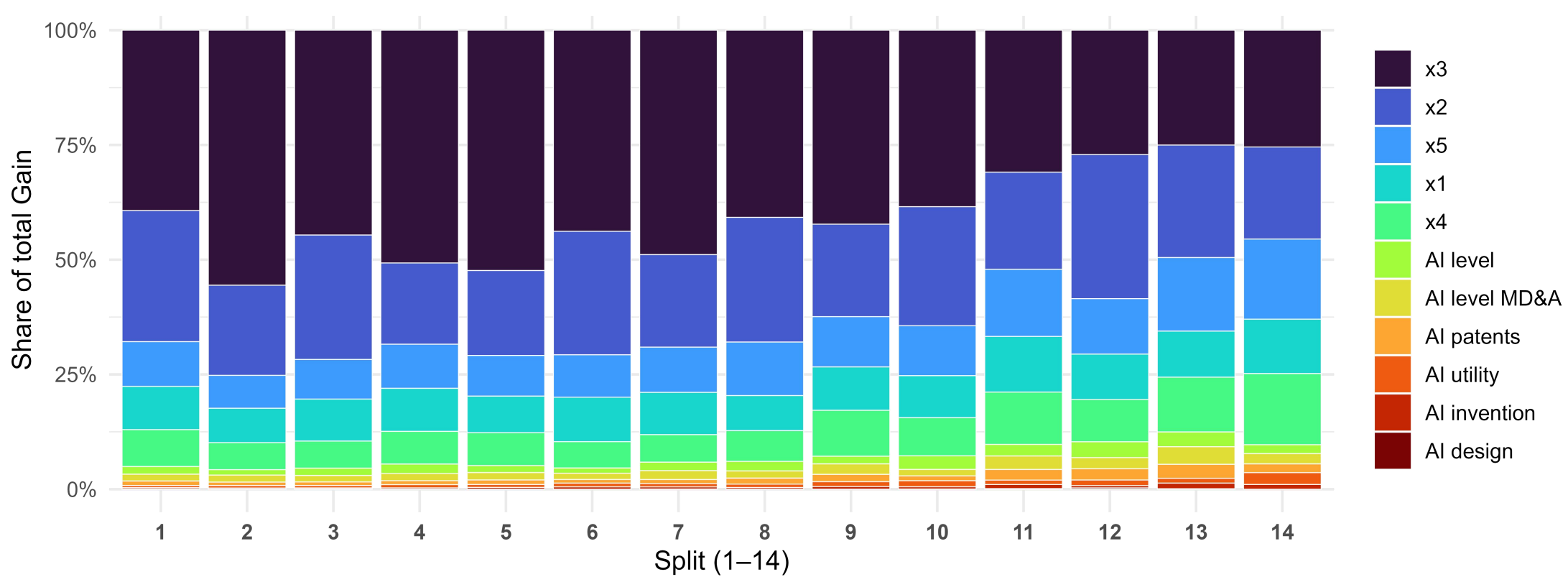

**Fig. 7** Stacked gain composition by split

Panel a of Figure 8 shows one distressed firm and how the AI signals tilt its score under different training windows. These local explanations illustrate how the model's AI signals can re-rank the same firm across regimes. For the distressed firm, the long and mid windows treat most AI proxies as reinforcing risk, so AI disclosure and AI patenting move in the same direction and the model assigns a very high score. In the short 2021 to 2022 window, the AI block diverges. Disclosure based AI signals pull against risk while the patent proxy pushes toward risk, and the net effect is a noticeably lower distress score for the same firm. For the healthy firm, the sign of the AI disclosure terms also shifts across windows, changing its distance to the classification boundary while remaining clearly low risk. Economically, this matches the broader evidence that the predictive meaning of AI variables is time varying. Under long training histories, AI measures align with higher predicted risk consistent with earlier regimes where they may correlate with transition uncertainty, while in the recent high adoption regime the model assigns more differentiated signs across disclosure and patent proxies.

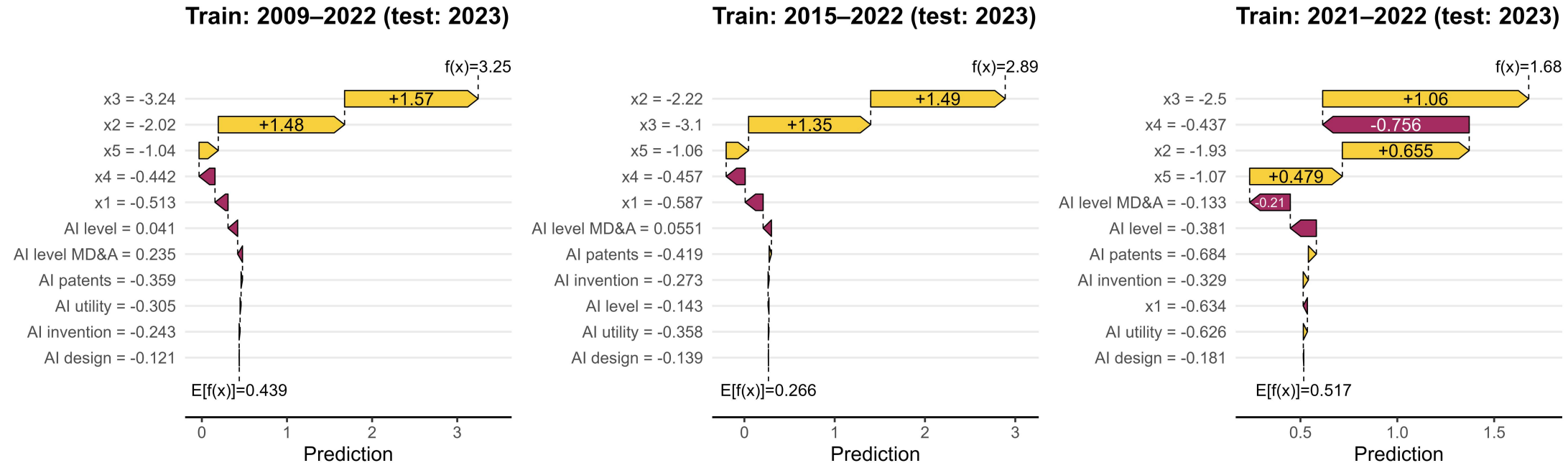

(a) Distressed company example

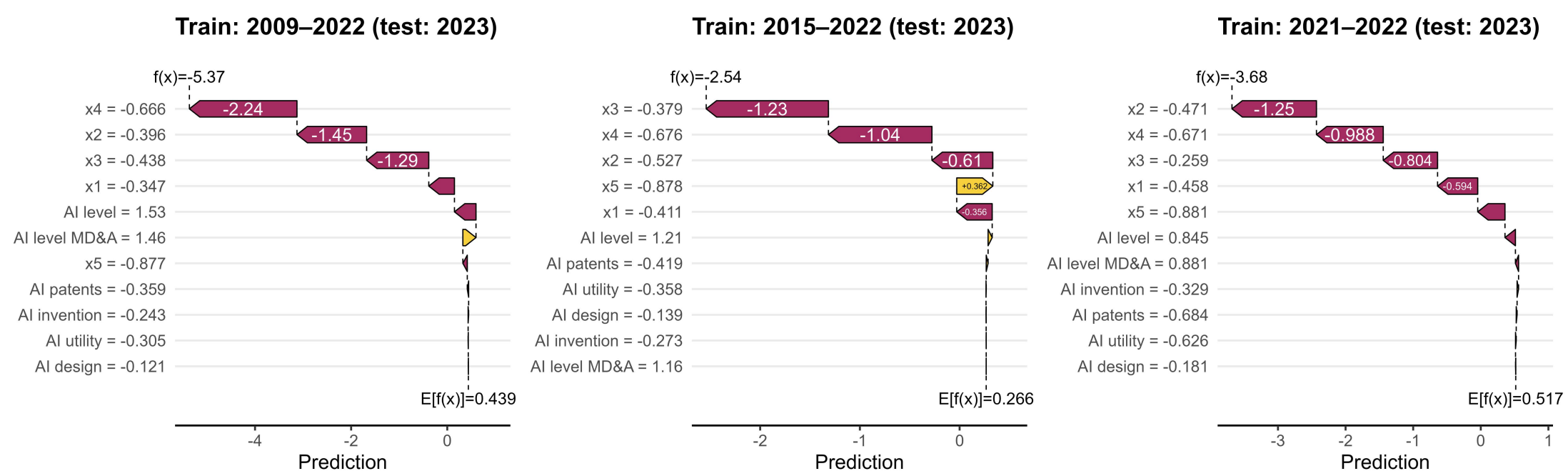

(b) Healthy company example

**Fig. 8** Local SHAP waterfall

Figure 9 shows that the sign reversals seen in the local waterfalls are not unique to those two firms. They line up with how the AI block behaves in the tails of the score distribution, and the pattern again depends on the training window. In the lowest risk tail, AI patent activity, especially LnInvention, is consistently protective and this pattern is most pronounced in the short 2021 to 2022 window. Economically, Total AI related invention output looks like an operational capability signal, so it supports a low default assignment for the safest firms. In the highest risk tail, the patent signal shifts direction and pushes toward distress, but its force is smaller than the disclosure based AI terms. In that same extreme tail, AI level and AI level MD&A become more influential and they push the score downward. Economically, among firms already flagged as risky, higher AI adoption and more detailed AI discussion can be read as active retooling and crisis response, so it modestly offsets distress risk even when patents still penalize weak, non durable innovation depth.

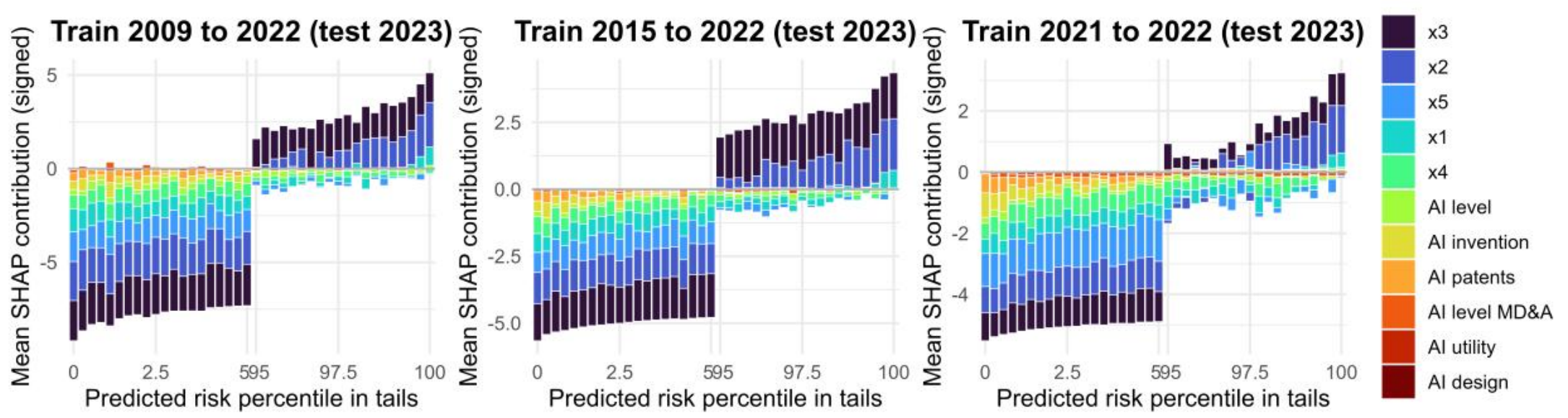

**Fig. 9** Mean signed SHAP by predicted risk tail percentiles

We also examined SHAP interaction patterns between the AI variables and the full set of financial predictors, and across AI variables themselves. Figure 10 illustrates a representative case, plotting the contribution of x3 across its range and colouring observations by AI level under each training window. The fitted response for x3 is stable and the colour gradient does not shift the curve in a systematic way. This same conclusion holds across the full interaction grid. We do not find meaningful evidence that AI measures amplify or dampen the effects of the financial ratios, or that the AI proxies interact strongly with each other. In economic terms, the AI variables enter the model as largely additive signals rather than as modifiers of the core financial channels.

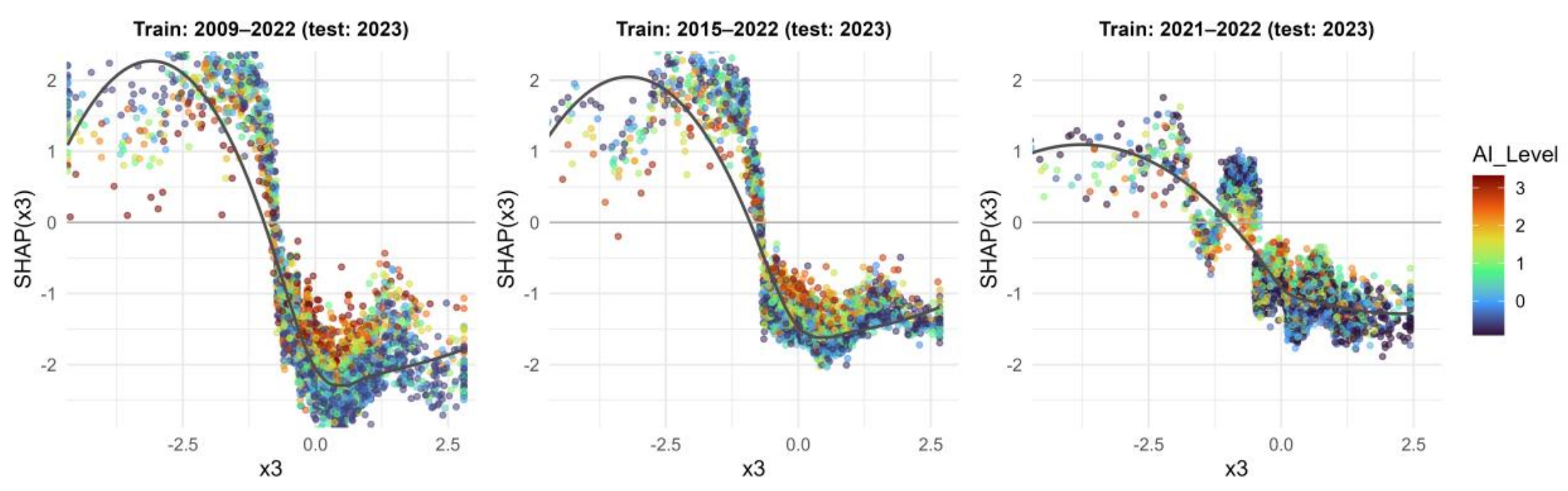

**Fig. 10** SHAP dependence plot for x3 with AI level colouring

Finally, while this study shows that AI-adoption signals add predictive power for forecasting financial distress in China's fast-evolving market, caution is warranted in generalizing the results. Part of the association may reflect that financially healthier firms are better able to fund the complementary intangible investments that AI enables and requires (Brynjolfsson et al., 2021). Yet such investment does not insulate firms from established failure drivers or macro-regulatory shocks (Fernández-Gámez et al., 2020; Sousa et al., 2022).

### 4.3 Robustness check

To ensure the validity and generalizability of our primary findings, we conduct robustness checks by replicating our methodology on alternative test years, 2021 and 2022, confirming that the incremental predictive value of AI-adoption metrics for early-warning signals persists across distinct economic and disclosure environments.

**Table 4**
Model Performance with AI Features on Alternative Test Years (2021 and 2022)

| | **2021** | | | | **2022** | | | |
|---|---|---|---|---|---|---|---|---|
| | **AUC** | **F1-Score** | **G-Mean** | **Recall** | **AUC** | **F1-Score** | **G-Mean** | **Recall** |
| XGB — with AI | **0.883** | **0.268** | **0.821** | 0.759 | **0.849** | **0.133** | **0.774** | **0.688** |
| XGB — without AI | 0.875 | 0.248 | 0.820 | 0.771 | 0.844 | 0.125 | 0.759 | 0.667 |
| LightGBM— with AI | **0.889** | 0.270 | **0.828** | **0.772** | **0.848** | 0.132 | **0.775** | **0.695** |
| LightGBM— without AI | 0.883 | 0.282 | 0.819 | 0.746 | 0.846 | 0.133 | 0.770 | 0.682 |
| RF— with AI | **0.894** | **0.276** | 0.830 | 0.773 | **0.835** | 0.122 | **0.774** | **0.701** |
| RF-— without AI | 0.893 | 0.272 | 0.837 | 0.790 | 0.833 | 0.126 | 0.772 | 0.694 |
| LR— with AI | **0.878** | **0.300** | 0.820 | 0.739 | 0.825 | **0.139** | 0.739 | 0.612 |
| LR— without AI | 0.872 | 0.275 | 0.827 | 0.768 | 0.830 | 0.134 | 0.743 | 0.624 |
| NN— with AI | 0.858 | 0.248 | 0.805 | 0.739 | 0.821 | **0.127** | 0.735 | 0.614 |
| NN— without AI | 0.878 | 0.277 | 0.821 | 0.755 | 0.832 | 0.124 | 0.739 | 0.627 |
| SVM— with AI | **0.895** | **0.316** | 0.815 | 0.724 | 0.832 | **0.132** | 0.756 | 0.653 |
| SVM— without AI | 0.880 | 0.288 | 0.832 | 0.770 | 0.852 | 0.127 | 0.760 | 0.665 |

Notes: This table employs the identical temporally-pruned window procedure described in Table 2, applied here to the 2021 and 2022 test years. The corresponding training windows are consequently shorter.

Table 4 shows that performance is materially weaker in 2022 than in 2021. AUC drops for every model family and F1 also falls sharply, indicating that 2022 represents a more challenging environment for distress prediction. Nevertheless, AI features continue to improve predictive performance in several model families, with the most reliable gains in tree based methods. This pattern aligns with Table 3, where improvements were strongest in discrimination and concentrated in XGBoost and Random Forest, with more limited evidence for other learners.

Figure 10 helps interpret these robustness results as distributional rather than deterministic. Several findings are consistent with the 2023 test results. Most notably, the final split, which relies on a single training year, is the least reliable, with visibly worse dispersion and sharp deterioration in recall oriented metrics, especially in 2022. By contrast, stronger realizations cluster in the three splits before the final one, in the shorter window region, and this pattern appears for both AI and no AI specifications. In those late splits, several runs achieve higher F1 and lower Type II error, reinforcing the practical relevance of recent windows for distressed firm identification. What does not carry over is the stronger 2023 result that the two year window is uniformly optimal, since the 2021 and 2022 test years do not exhibit a single dominant window choice.

Even so, the AI specification tends to dominate the baseline on F1, especially in the shorter window region. This implies a more favorable predictive balance between capturing distress cases and limiting false alarms, which is central for early warning screening. The advantage is visible in both split averages and several late split realizations where the AI dots lie well above the no AI dots. The picture is less clear for AUC. In 2021, many learners show higher AUC with AI, but a few weak runs reduce the average gap. In 2022, there is no systematic AUC advantage, although some AI runs still appear at the top of the distribution.

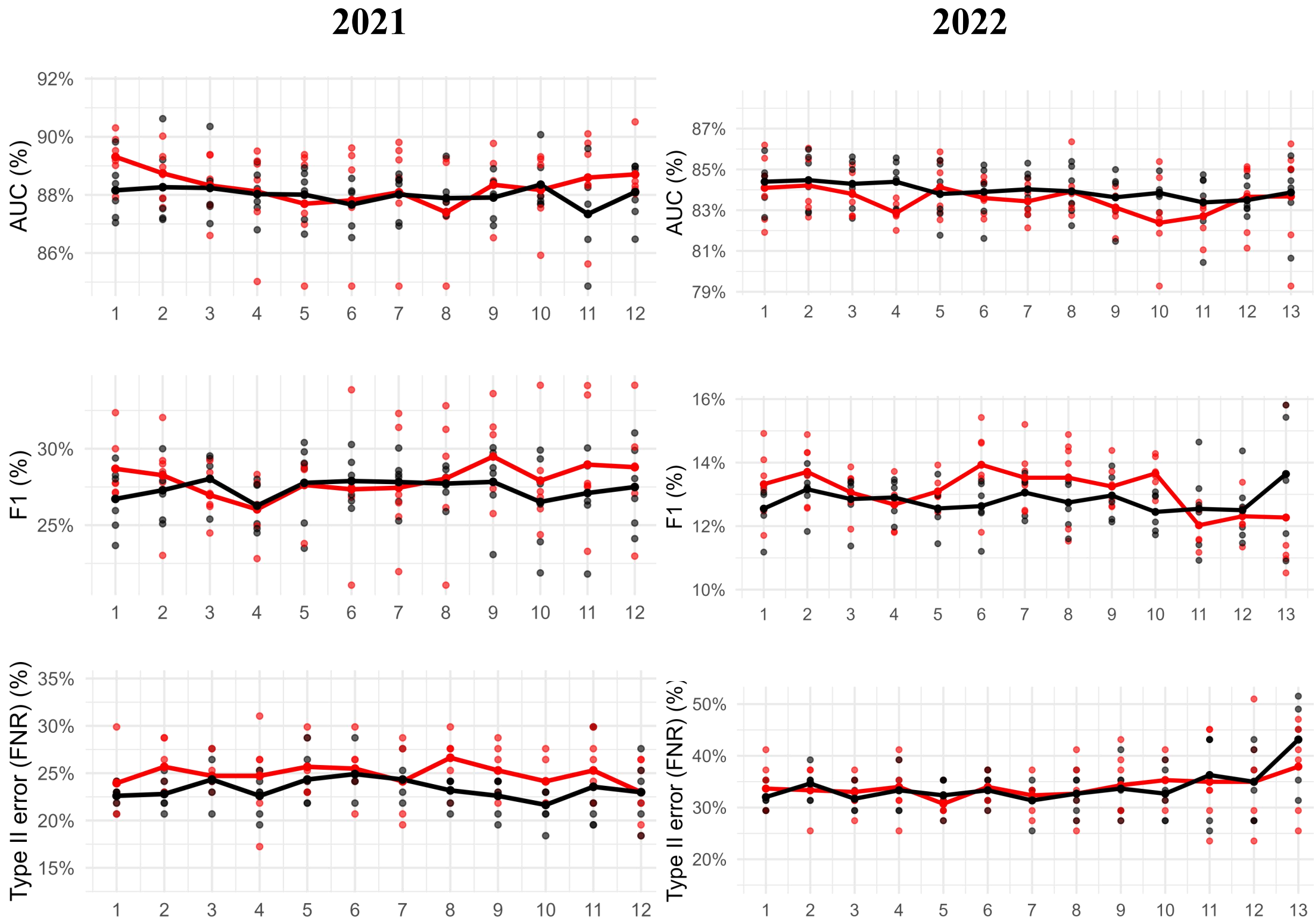

**Fig. 11** Split wise model performance (2021 and 2022 test).

### 4.4 Generalizability and the Chinese Institutional Setting

The findings of this study demonstrate that AI-adoption signals provide significant incremental power for forecasting financial distress within China's dynamic market. However, the generalizability of these results to other institutional contexts requires careful consideration. The Chinese setting is characterized by a distinct, policy-driven model of technological diffusion, heavily influenced by top-down initiatives. The 2017 "New Generation Artificial Intelligence Development Plan" is not merely a strategic document but a comprehensive national blueprint that set the explicit goal of making China the world leader in AI by 2030, monetizing it into a trillion-yuan industry, and shaping global ethical standards (Roberts et al., 2021). This stands in contrast to the more market-led adoption patterns often observed in liberal economies.

The implementation of this strategy creates a unique ecosystem. The central government provides a "stamp of approval" and incentivizes action, but the actual innovation and transformation are driven by local governments and a selected "national team" of private champions like Baidu, Alibaba, and Tencent, which receive preferential treatment to focus on strategic AI sectors (Roberts et al., 2021). This state-guided, yet decentralized, execution model, combined with political incentives for local officials to fulfill central government initiatives, creates a powerful engine for policy-driven adoption that may differ fundamentally from corporate AI investment decisions in other countries (Khanal et al., 2024).

Consequently, the predictive relationship we identify may be shaped by these unique institutional factors, including the structure of state-firm relationships and specific disclosure incentives within China (Chen et al., 2024; Dang and Motohashi, 2015). The AI metrics captured in our study likely reflect not only genuine technological integration but also strategic responses to this distinct policy landscape. Therefore, the external validity of our specific coefficient estimates may be limited, as the observed relationship between AI disclosure and financial distress is embedded within China's particular socio-political and economic context.

## 5. Conclusion

This study shows that firm level AI adoption indicators from corporate disclosures and patent filings provide statistically and economically meaningful incremental predictive power for forecasting financial distress among Chinese non financial listed firms. Using six machine learning classifiers and a temporally pruned training window design, we find that adding AI features to financial features improves out of sample classification, with the most robust gains in tree based ensembles, especially XGBoost and random forests. Improvements in AUC, F1 score, and G Mean are driven mainly by lower Type II error, meaning fewer missed distressed firms in an early warning context.

Our methodological contribution also concerns how historical data should be used in dynamic prediction. Performance is non monotonic in window length, so longer histories do not automatically improve forecasts. Models trained on a single year are unstable, while windows focused on recent regimes often deliver stronger distress identification, particularly when AI signals are included. This highlights the practical importance of aligning the training window with the prediction environment under evolving regimes. The pruned window framework preserves chronological validity while keeping models responsive to current conditions.

Explainability results clarify why AI proxies exhibit mixed signs without undermining predictive coherence. SHAP shows that financial fundamentals remain the primary contributors to predicted distress, while AI variables add incremental signal in high adoption regimes captured by shorter windows. Higher AI disclosure in annual reports and higher invention oriented patent activity are more often associated with lower predicted distress, while higher MD&A AI disclosure and design patent activity are more often associated with higher predicted distress, with directions varying across regimes. These patterns are economically interpretable as correlational signals rather than structural effects, reflecting that AI proxies can capture capability in some settings and transition strain or strategic signaling in others.

These findings have practical relevance for lenders, regulators, exchanges, and internal risk teams using early warning systems. First, temporal validation is critical, as models that perform well in pooled samples may be fragile near the decision margin. Second, training window governance is a low cost but high impact design choice, since recent rolling or pruned windows can materially improve detection without changing model class. Third, AI related disclosure and patent proxies are operationally useful leading indicators that update faster than realized financial distress and should be interpreted as predictive risk signals, with ongoing drift monitoring as disclosure norms evolve.

The study also has limitations. AI measures are proxies based on disclosures and patents, blending adoption with reporting and policy incentives. The Chinese institutional setting and the ST based distress definition may limit external comparability. SHAP improves transparency but does not identify mechanisms, and the objective remains robust prediction under drift rather than causal inference. Several extensions are natural. Future work can separate talk from capability using richer text methods, including contextual embeddings and topic models. Patent measures can be refined using quality indicators such as citations and renewals, and window based learning can be integrated with domain adaptation. Decision focused evaluation that assigns explicit costs to missed distress and false alarms would further align prediction with economic monitoring objectives.

**Declaration of generative AI and AI-assisted technologies in the manuscript preparation process**

During the preparation of this work, the author(s) used ChatGPT for language polishing and stylistic refinement. After using this tool, the author(s) carefully reviewed and edited the content as needed and take full responsibility for the accuracy, originality, and integrity of the published article.